\documentstyle[12pt]{article}
\begin{document}
\renewcommand{\thesection}{\arabic{section}.}
\newcommand{\tq}{top quark }
\newcommand{\nn}{\nonumber}
\newcommand{\nl}{\newline}
\newcommand{\gp}{g^{\prime} }
\newcommand{\gpp}{g^{\prime\prime} }
\newcommand{\be}{\begin{equation}}
\newcommand{\ee}{\end{equation}}
\newcommand{\bea}{\begin{eqnarray}}
\newcommand{\eea}{\end{eqnarray}}
\newcommand{\lb}{\label}
\newcommand{\pr}{\prime}
\newcommand{\al}{\alpha}
\newcommand{\ga}{\gamma}
\newcommand{\de}{\delta}
\newcommand{\De}{\Delta}
\newcommand{\eps}{\epsilon}
\newcommand{\elw}{ electroweak}
\newcommand{\ti}{\times}
\newcommand{\half}{\frac{1}{2}}
\newcommand{\ra}{\rightarrow}
\newcommand{\lra}{\longrightarrow}
\newcommand{\lp}{\Lambda_P}
\newcommand{\mi}{\mbox{i}}
\newcommand{\mtr}{\mbox{Tr}\,}
\newcommand{\ttttil}{\tilde{T}}
\newcommand{\ttpmtil}{\tilde{T}_{\pm}}
\newcommand{\ttztil}{\tilde{T}_{0}}
\newcommand{\ttptil}{\tilde{T}_{+}}
\newcommand{\ttmtil}{\tilde{T}_{-}}
\newcommand{\tytil}{\tilde{Y}}
\newcommand{\ttt}{\bar{T}}
\newcommand{\ttpm}{\bar{T}_{\pm}}
\newcommand{\ttz}{\bar{T}_{0}}
\newcommand{\tti}{\bar{T}_{i}}
\newcommand{\ttj}{\bar{T}_{j}}
\newcommand{\ttk}{\bar{T}_{k}}
\newcommand{\ttdrei}{\bar{T}_{3}}
\newcommand{\ttzwei}{\bar{T}_{2}}
\newcommand{\tteins}{\bar{T}_{1}}
\newcommand{\ttdreitil}{\tilde{T}_{3}}
\newcommand{\ttzweitil}{\tilde{T}_{2}}
\newcommand{\tteinstil}{\tilde{T}_{1}}
\newcommand{\ttp}{\bar{T}_{+}}
\newcommand{\ttm}{\bar{T}_{-}}
\newcommand{\ty}{\bar{Y}}
\newcommand{\tgr}{\bar{g}_R}
\newcommand{\dd}{1-2\De}
\newcommand{\lr}{SU(2)_L \ti SU(2)_R \ti U(1)_{B-L}}
\newcommand{\sm}{SU(2)_L \ti U(1)_Y}
\newcommand{\sutq}{SU(2)_q}
\renewcommand{\thefootnote}{\fnsymbol{footnote}}
\begin{titlepage}
\begin{quote}
\raggedleft   . \end{quote}
\vspace{2cm}\begin{center}
{\bf
Electroweak Parity Breaking by Isospin Deformation} \\ {\bf
and the Copenhagen Vacuum}\footnote{
To appear in the proceedings of 28th
Symposion on Elementary Particle Physics, Wendisch-Rietz,
August 30 - September 3, 1994}
\end{center}
\vspace{3cm}
\begin{center} {R. B\"onisch}
\\
DESY-IfH, Platanenallee 6, D-15738 Zeuthen
\end{center}\vspace{3cm}
\begin{abstract}
{}From the global chiral $SU(2) \times U(1)$ symmetry that
appears in the electroweak Left-Right Model in the fundamental representation,
a continuous transition to the representation of the
Minimal Standard Model is considered in the Cartan subalgebra of the
right-handed sector. The connection parameter $\De$ is the splitting of
$U(1)_R$ quantum numbers. $\De$ is a deformation parameter, breaks
$SU(2)_R$ and parity and is proportional to an isomagnetic field, leading
to a Copenhagen vacuum structure. A simple mapping on the fundamental
representation of $SU(2)_q$ gives $\De$
in terms of $q$.
\end {abstract}
\end{titlepage}
\renewcommand{\thefootnote}{\fnsymbol{footnote}}
Generalizations of Lie algebras which connect various classical
symmetries with one another
have been known in mathematics as expanions \cite{Gilmore}. In physics they
are used to account for deviations from a classical Lie symmetric
structure. Such generalizations
are particularly useful when generalized generators can be written
with an inbuilt classical limit of the Lie algebraic representation.
This is the case for the $q$-deformation $SU(2)_q$ of $SU(2)$
\cite{rep}. Strictly speaking, the generalization parameter $q$ is a symmetry
breaking parameter. In the Heisenberg spin model for example, $q$ appears
as the anisotropy of an $XXZ$ interaction \cite{Ruiz-Altaba}.
Deformed spinors
are crucial in the construction of momentum vectors with a discrete
set of eigenvalues, such
that $q \neq 1$ breaks the continuous space-time symmetries in a known
and
controlled way \cite{Wess}. The resulting non-commutative geometry is
expected
to
yield finite theories \cite{Zoupanos,Kempf} and can therefore be regarded
as a new regularisation science,
which might make contact with other schemes like e.g. FUT's \cite{fut},
string theory, where
effective uncertainty relations among positions occur \cite{string}, or
the Connes-Lott version of the Standard Model \cite{Connes} one day.
The latter approaches make definite use of the interplay between
internal symmetries and external geometry within the introduction of
a fundamental scale parameter.
Phenomenologically well motivated
electroweak models, where a composite Higgs regulates
the Fermi theory and on the other hand is used to look for possible
internal symmetries beyond the SM \cite{cond},
should actually be embedded in such a geometrically more advanced scheme.
As a matter of taste, it might nevertheless be taken as an
advantage to ask experiment for any possible information first.

In this sense it might be worthwhile to reconsider the breaking of
space-time parity by
electroweak interactions using generalized algebras, like it was
presented in \cite{Bonisch}.
To this end we define a transition from the minimal SM with $\sm$
symmetry to the left-right (LR) model with $\lr$. To be precise, the
transition connects the generators of the
Cartan subalgebras of the global symmetries.
The generalization or transition parameter $\De$
stands for explicit breaking of parity and (by deformation) of $SU(2)_R$.
The fundamental representation of $\sutq$ is, up to a gauge rotation,
identical to that one of $SU(2)_\De$,
giving a very simple expression of $\De$ in terms of $q$.
Some statements on the special generalized local symmetry are immediate:
A $U(1)_{\ttz}$
subsymmetry of $SU(2)_R$ that is preserved is given by $\De$ and anomalies
are fixed to cancel in the remaining $U(1)_{\ttz} \ti U(1)_{\ty}$,
while the topology
of $SU(2)_\De$ can be required to be unchanged from the classical limit
and anomalies should neither appear there.
The present physical situation is infact just a gauge model in an
chromomagnetic, or isomagnetic background, yielding the well-known
Copenhagen Higgs Lagrangian \cite{Savvidy,Nielsen-Olesen,stable}.

The kinetic and interaction Lagrangian is
\be {\cal L}_x = \mi \bar{\Psi}_x \ga_\mu D^\mu_x \Psi_x, \lb{ia}\ee
where $x$ is summed over $L,R$; $\Psi$ is an isospinor doublet,
\be D^\mu_x = \partial^\mu + \mi g_x \frac{{\bf T}_x}{2} {\bf W}^\mu_x
+ \mi g^\pr Y_x B^\mu, \lb{ia2}\ee
$T_x$ are the generators of $SU(2)_x$, in the fundamental representation
given by
Pauli matrices $\tau_i$ (normalized to $\tau^2 =1$) as
$T_0=\frac{1}{2} \tau_3$ and
$T_\pm = \frac{1}{\sqrt{2}} (\tau_1 \pm \mi \tau_2)$,
${\bf W}_x$ the corresponding
gauge field triplets and $B$ the $U(1)_Y$ gauge field which couples
to vectorlike currents only. We need not
consider any spontaneously broken phases, nor family or color degrees of
freedom for our purposes here. There is an underlying global chiral
symmetry
\be G = SU(2)_L \times SU(2)_R \times U(1)_{Y_L}  \ti U(1)_{Y_R}\lb{gg}\ee
and the electric charge operator is defined as
\be Q_x = T_{0x} + Y_x, \lb{q}\ee
in each helicity projection
of $J_{em}=\bar{\Psi} \ga^\mu Q \Psi$ such that \- $Q_L =Q_R=Q
=$\- diag\-$(0,-1)$
\- [diag \- $(2/3, -1/3)]$ for lepton- [quark-] doublets.

In both the LR and the SM model, $SU(2)_L$ is in the fundamental
representation, and \be  Y_L=\frac{B-L}{2}, \ee
where $B=1/6$ is the baryon number and $L=1/2$ the lepton number.
In the LR model \cite{Senjanovic+Mohapatra},
\be {\bf T}_L={\bf T}_R,
\hspace{1.8cm}  Y_L= Y_R.
\lb{lrn}\ee
Parity is defined by
\be P: {\bf x} = (x_0, x_j) \ra {\bf x}^\pr = (x_0, -x_j) \ee
for coordinates,
\be \psi ({\bf x}) \ra \psi^\pr({\bf x}^\pr) = \ga_0 \psi ({\bf x}^\pr) \ee
for spinors and
\be G_\mu({\bf x})  \ra G_\mu^\pr({\bf x}^\pr) = \eps(\mu)G_\mu({\bf x}^\pr)\ee
for gauge fields $G_\mu$, where $\eps(0)=1$ and $\eps(j)=-1$.
With the LR assigment eq. (\ref{lrn}),
the Lagrangian eq. (\ref{ia}) is invariant with respect to $P$ if $g_L=g_R$
and the breakdown will have to occur in low energy states only.

In the SM  we have
\be {\bf T}_R = 0,  \hspace{3cm} Y_R = Q \lb{smn}\ee
instead of the $R$-numbers in eq. (\ref{lrn}).
No preserved parity exists for this hidden $SU(2)_R$ \cite{Grimus},
$P$ is broken explicitely in the non-abelian as well as in the abelian
primordial gauge sector.
That nevertheless no breaking term appears explicitely in $\cal L$ is somewhat
unsatisfactory, therefore we make an attempt to set such a term free:
Using the splitting of hypercharge quantum numbers of
(potential) isospin components, eq. (\ref{smn}),
\be \Delta \equiv \frac{y^u_R - y^d_R}{2}, \lb{delta}\ee
as a continuous parameter $0 \leq \De \leq 1/2$, we can get
to eq. (\ref{smn}) from eq. (\ref{lrn}) by mapping
the generators $(Y_R, T_{0R})$ of the Cartan subalgebra $C_{V}$ on
generalized quantities
\bea \left( \begin{array}{c} \ty \\ \ttz \end{array} \right)
= \left( \begin{array}{cl} 1 & 2\De \\ 0 & 1\!\!-\!\!2\De \end{array} \right)
\left( \begin{array}{c} Y_R \\ T_{0R} \end{array} \right).
\lb{bmix}\eea
For $\De \neq 0$, isospin components have different $U(1)$ charges and break
\be SU(2)_R \stackrel{\De \neq 0}{\lra} U(1)_{\ttz} \lb{break}\ee
as is seen by redefining $SU(2)_R$ elements
$g_R \ra \bar{g}_R(\De) \equiv (1\!-\!2\De)g_R$ and
probing the generalized hypercharge interaction term to transform like
\be \mi \bar{\Psi}_R \ga^\mu \ty \Psi_R B_\mu
\stackrel{{\bf T}_R}{\longrightarrow}
\mi \bar{\Psi}_R \ga^\mu \ty \Psi_R B_\mu
- \tgr \bar{\Psi}_R \ga^\mu [\ty, T_j\omega^j] \Psi_R B_\mu. \ee
The $T_0$ part in $\ty$ is nonabelian,
\be [\ty, T_0]=0, \hspace{2cm} [\ty, T_\pm]=\pm 2\De T_\pm \lb{tybar}\ee
such that the breaking is proportional to
$\tgr \cdot \De$, which vanishes in the LR representation at
$\De =0$ and in the SM representation at $\De = 1/2$, where the action of
$SU(2)_R$ becomes trivial. Here, $\ty$ violates $SU(2)_R$ just like
Yukawa couplings $g_{top} \neq g_{bottom}$
violate the custodial $SU(2)$ in the SM scalar sector.
The transition eq. (\ref{bmix}) preserves eq. (\ref{q}) and $\ty$ is always
traceless so that no axial anomalies infect the $U(1)_{\ttz} \ti
U(1)_{\ty}$ subsector.

Instead of redefining all $SU(2)_R$ elements it is straightforward to
take the map eq. (\ref{bmix}) as a deformation on the fundamental
representation, following the work of Curtright and Zachos
on various kinds of deformations \cite{rep}.
With $\ttz$ and $T_\pm$ we have $\De$ in the commutators,
\be  [\ttz, T_\pm] =\pm  (1\!-\!2\De) \,\, T_\pm,
\hspace{1.2cm} [T_+,\, T_-]= (1\!-\!2\De)^{-1} \ttz
\lb{quom}\ee
or, for the real representation,
\be [ \tti, \, \ttj] = \mi \bar{\eps}_{ijk} \ttk, \lb{quom2}\ee
\be \bar{\eps}_{ijk} = \left\{ \begin{array}{ll}
\pm (1\!-\!2\De) ^{-1}  & \mbox{for $i\,(j)=1\,(2),$  $k=3$} \\
\pm (1\!-\!2\De) &  \mbox{for $i$ or $j=3,$  $k=1$ or $2$} \\
0 & \mbox{else} \end{array} \right. \lb{bareps}\ee
where $\bar{\eps}_{ijk}$ becomes the ordinary $\eps_{ijk}$ when $\De \ra 0$
and $T_\pm = \frac{1}{\sqrt{2}}(\tteins \pm \ttzwei)$.
The Casimir may be written as
\be C_\De = 2 \,T_+\, T_- + \ttz \left[\ttz-(1\!-\!2\De)^{-1}\right]
\lb{casi}\ee and is minimal at $\De = 1/2$.
The corresponding classical breaking term is the rescaling of the 3rd
direction or anisotropy proportional $\De$:
For the Lagrangian of
a classical free particle, the deforming map eq. (\ref{bmix}) gives
\bea L &\ra& L^\pr = \frac{1}{2m}p^{\pr2} = L - U \lb{lu}\\
U&=&-\frac{2\De}{m}p_3^2 +..., \eea
where $U= - {\bf A} \cdot{\bf v}$ contains a vector potential
\be {\bf A} \equiv (0,\,0,\,A_3=2 \De p_3).\lb{a}\ee
The breaking term is thus a magnetic field
${\bf H} = \nabla \ti {\bf A}$, which causes the breakdown eq. (\ref{break})
and we are dealing with some kind of `Zeeman-deformation'.
The SM quantum numbers eq. (\ref{smn}) are reached at $\De \ra 1/2$, where
${\bf A}$ diverges.

Iso- and
chromomagnetic backgrounds have been studied as models for the QCD
vacuum
already in the seventies, because the minimum energy density appears to
be at finite $H \sim \Lambda^2$ \cite{Savvidy}. An imaginary contribution
to the energy density represents a massive harmonic oscillator mode,
which is lower in energy than the static field
and behaves like a 1+1 dimensional tachyonic Higgs mechanism
\cite{Nielsen-Olesen}. Stability conditions have been investigated
\cite{stable}.
A lattice study with anisotropically shifted link variables was done in
\cite{lattice}.
Recent results have been reported by K.-J. Biebl and H.-J. Kaiser at the
meeting \cite{proc}.
A condensation enhancement in the 2+1 dimensional
Nambu-Jona-Lasinio model in magnetic backgrounds was recently demonstrated
\cite{njl}.

A factor analogous to
the ones in the commutation relations eq. (\ref{quom}) and
(\ref{quom2}) appears in the fundamental representation of the much discussed
$q$-deformation $\sutq$ of  $SU(2)$ in the form
\be [\ttz,\,\ttpm]=\pm \ttpm, \hspace{2cm}
[ \ttp, \ttm]= [\ttz]_{q^2}, \lb{su2q}\ee
where $[x]_{q^2} \equiv (q^{2x}-q^{-2x}/(q^2-q^{-2})$.

The solution of eq. (\ref{su2q}) $\bar{T}$ in terms of the classical
$T$'s is
\bea
\ttz &=& T_0,\nn \\
\ttp &=&  \sqrt{ \frac{2}{q+1/q} \cdot
\frac{[T+T_0]_q\,[1+T-T_0]_q}{(T+T_0)\,(1+T-T_0)}} \, T_+, \nn \\
\ttm &=&  \sqrt{ \frac{2}{q+1/q} \cdot
\frac{[T-T_0]_q\,[1+T+T_0]_q}{(T-T_0)\,(1+T+T_0)}} \, T_-.\lb{ttbar}\eea
For $T=1/2$ we get
\be \ttpm=\sqrt{2/(q+1/q)} T_\pm \lb{ttpmq}\ee
and $H$ is in the 3rd cartesian direction,
\be [\tteins,\,\ttzwei]=\mi\al^2 T_3,
\hspace{.5cm} [T_3,\,\tteins]=\mi \ttzwei,
\hspace{.5cm} [\ttzwei,\,T_3]=\mi \tteins, \ee
where $\al=\sqrt{2/(q+1/q)}$.
Rotating ${\bf \hat{e}}=R \,{\bf e}$,
\be R= \left( \begin{array}{ccc} r^2 & -r^2 & -r \\ -r^2 & r^2 & -r \\
r & r& 0 \end{array} \right), \lb{r}\ee
where
$r= \sin (\pi/4)$, reshuffles the factor and after absorbing the potential
into the coordinates by
${\bf e} \ra {\bf e}^\pr =(\al\,e_1,\al\,e_2,\,e_3)$
we get new coordinates
\bea
\hat{e}_1^\pr &=& \al r^2 e_1 - \al r^2 e_2 - r e_3 \nn \\
\hat{e}_2^\pr &=& -\al r^2 e_1 + \al r^2 e_2 - r e_3 \nn \\
\hat{e}_3^\pr &=& \al r (e_1 + e_2) = \al (\hat{e}_1 + \hat{e}_2).\eea
With
\be 1-2\De=\sqrt{\frac{2}{q+1/q}}, \lb{conn}\ee
$\hat{e}_3^\pr$ fulfills the central requirement eq. (\ref{bmix}).
The limits $q \ra 1$ and $q \ra \infty$ are recovered by $\De \ra 0$
and $\De \ra \half$ in eq.  (\ref{conn}).
Martin-Delgado has previously interpreted the deformation as an
$H$-field effect \cite{Martin-Delgado}. There
the symmetry $q \leftrightarrow q^{-1}$
corresponds to a freedom in the sign of $H$.

All above solutions of commutation relations use finitenes of the
representation, i.e. the requirement of highest and lowest weight states
to
exist for making the difference constant vanish. The latter would otherwise
appear as a constraint in classical equations of motion.
{}From eq. (\ref{ttbar}) as well as in eq. (\ref{bmix}) and (\ref{quom}),
the generalized generators $\ttpm$
create the vacuum like in the classical limit,
$\ttpm
|t,t_0 \pm 1 \rangle =0$, and shall thus preserve the topology and avoid
singularities in the functional measure of the generalized action.
\vspace{.5cm}\nl
{\bf Acknowledgement}\nl
I wish to thank the organizers for the invitation and very nice conference.
For a number of helpful discussions I thank K.-J. Biebl and G. Zoupanos.

\end{document}